\documentstyle[12pt]{article}
\input{epsf}
\newcommand{\be}{\begin{eqnarray}}
\newcommand{\ee}{\end{eqnarray}}
\newcommand{\ba}{\begin{array}}
\newcommand{\ea}{\end{array}}

\begin{document}

\rightline{\textsl{\date{\today}}} \vspace{0.5cm}
\begin{center}
{\Large On Narrow Nucleon Excitation N$^*$(1685)}\\
\vspace{0.35cm}
 V. Kuznetsov$^{1,2}$ , M.V. Polyakov$^{3,4}$ and M. Th\"urmann$^4$\\

\vspace{0.35cm}
$^1$ Kyungpook National University, 702-701, Daegu, Republic of
Korea\\
$^2$ Institute for Nuclear Research, 117312, Moscow, Russia, \\
$^3$Petersburg Nuclear Physics
Institute, Gatchina, St.\ Petersburg 188300, Russia\\
$^4$Institut f\"ur Theoretische Physik II, Ruhr--Universit\"at
Bochum, D--44780 Bochum, Germany
\end{center}

\begin{abstract}
\noindent
We collected notes and simple estimates about putative narrow nucleon N$^*$(1685)-- the
candidate for the non-strange member of the exotic anti-decuplet of baryons. In particular, we
consider the recent high precision data on $\eta$ photoproduction off free proton obtained by
the  Crystal Ball Collaboration at MAMI. We show that 
it is difficult to describe peculiarities of
these new data in the invariant energy interval of $W\sim 1650-1750$~MeV in terms
of known wide resonances. Using very simple estimates, we show that the data
may indicate an existence of a narrow N$^*(1685)$ with small photocoupling to the proton.
\end{abstract}

%\PACS{13.60.Le\and14.20.Gk}

\section*{Introduction to the neutron anomaly}

The prediction of light and narrow anti-decuplet of baryons in the framework of the chiral quark soliton model ($\chi$QSM) \cite{dia}
has a direct implication for the classical field of nucleon resonances spectroscopy:
one should expect an existence of the nucleon state, which is much narrower than the usual nucleon excitations with analogous 
mass~\cite{dia,dia1,arndt,michal}.

An important observation was made  in Ref.~\cite{max},  it was demonstrated that the nucleon resonance from the anti-decuplet has a clear
imprint of its exotic nature: the anti-decuplet nucleon is excited predominantly by the photon  from the neutron, its photoexcitation
from the proton target is strongly suppressed. Therefore the $\gamma n\to\eta n$ process has been suggested
in Ref.~\cite{max} as a ``golden channel"  to search for the anti-decuplet nucleon.
The modified partial wave analysis (PWA) of the elastic $\pi N$ scattering \cite{arndt} showed that the existing data on $\pi N$
scattering can tolerate a narrow $P_{11}$ resonance if its $\pi N$ partial decay width is below $0.5$~MeV and it has the mass
around 1680~MeV.

Recently four groups - GRAAL~\cite{gra0,gra1},
CBELSA/TAPS \cite{kru}, LNS~\cite{kas}, and Crystal
Ball/TAPS~\cite{wert} - reported an evidence for a narrow structure
at $W\sim 1680$~MeV in the $\eta$ photoproduction on the neutron.
The structure was observed as a bump in the quasi-free cross
section (the neutron anomaly\footnote{The name ``neutron anomaly" was introduced in Ref.~\cite{jetp} to denote
the bump in the quasi-free $\gamma n\to \eta n$ cross section around $W\sim 1680$~MeV and its absence in the quasi-free $\gamma p\to \eta p$ cross section.})
and as a peak in the invariant-mass spectrum of the
final-state $\eta$ and the neutron ($M(\eta n)$)
~\cite{gra1,kru,wert}. The width of the bump in the quasi-free
cross section is close to that expected due to the smearing of the
target neutron bound in a deuteron target by Fermi motion. The
width of the peaks observed in the $M(\eta n)$ spectra is close
to the instrumental resolution of the corresponding
experiments~\cite{gra1,kru,wert}.

Furthermore, a sharp resonant structure at $W\sim 1685$~MeV was
found in the GRAAL data on the beam asymmetry for the $\eta$
photoproduction on the free proton ~\cite{acta,jetp}. Such
structure is not (or poorly) seen in the $\gamma p \to \eta p$
cross section~\cite{crede,cr2}. 

In Refs.~\cite{gra1,acta,jetp,az,kim,tia}, the combination of the
experimental findings was interpreted as a signal of a nucleon
resonance with the mass near $\sim 1680$~MeV and unusual properties:  the
narrow width and the stronger photoexcitation on the neutron comparing to 
that on the proton.
Alternatively, the authors of  Refs.~\cite{ani,skl} explained
the neutron anomaly in terms of the
interference of well-known resonances and in Ref.~\cite{dor} due to effects of meson loops\footnote{It is worth noting
here that the models of Refs.~\cite{ani,skl,dor} do not predict the neutron anomaly, the observed peak in the
neutron cross-section (and its apparent absence in the proton channel) has been used as an input for fitting of 
quite numerous model parameters.}.

\noindent
In year 2010 more results on the neutron anomaly were obtained:
\begin{itemize}
\item
In Ref.~\cite{Compton}
the first study of quasi-free Compton scattering on the neutron in
the energy range of $E_{\gamma}=750 - 1500$~MeV was performed. The
data reveal a narrow peak at $W\sim 1685$~MeV.
Such peak is absent in the Compton scattering on the proton as well as in the reactions
$\gamma n\to \pi^0 n$ and $\gamma p\to \pi^0 p$. The latter observation implies that
the putative narrow resonance should have very small $\pi N$ partial width, that is in agreement 
with modified PWA of Ref.~\cite{arndt} and with theoretical expectations for the nucleons from the 
anti-decuplet \cite{dia,dia1,arndt,michal}.  
In other words, the neutron anomaly was also observed in the
Compton scattering. For details of the corresponding analysis see Ref.~\cite{Compton}.

 We note that the explanations of the neutron anomaly in the $\eta$ photoproduction
in terms of the
interference of well-known resonances \cite{ani,skl}  and  due to effects of meson loops~\cite{dor}
obviously do not work  in the case of the Compton scattering.

\item
 Recently the data of the CBELSA/TAPS  collaboration \cite{kru} on $\eta$ photoproduction off the neutron have been
reanalysed by the same collaboration. Namely, the de-folding of the Fermi motion has been performed.
The corresponding preliminary results were presented by B.~Krusche at MESON10 workshop in Krakow \cite{krusche}.
One can use the results of this new analysis in order to extract the photocoupling of neutral component
of N$^*$(1685). The method is described in Ref.~\cite{az}, following it one can easily obtain:

\begin{eqnarray}
\label{an}
\sqrt{{\rm Br}_{\eta N}} A_{1/2}^n \sim 15\cdot 10^{-3}\ {\rm GeV}^{-1/2} \ \ \ {\rm (CBELSA/TAPS\ data)}
\end{eqnarray}
That value of the photocoupling is in a striking agreement with the value obtained
in Ref.~\cite{az} from the analysis of the GRAAL data of Refs.~\cite{gra0,gra1}.

\item
The neutron anomaly is also seen in $\eta$ photoproduction on $^3$He, see preliminary data of the A2 collaboration
in Master Theses of L.~Witthauer \cite{witthauer}. The position of the bump in neutron quasi-free cross section
is in agreement with the position of the corresponding bump obtained from the deuteron scattering in Refs.~\cite{gra0,gra1,kru,kas,wert}.
The width of the bump in $^3$He is larger than extracted from deuteron scattering, that is due to more wider Fermi momentum distribution
in $^3$He nucleus.

Observation of the neutron anomaly in the scattering on new type of the nuclei ($^3$He) is important in order to exclude
an appearance of the neutron anomaly due to nuclear and/or rescattering effects.

\end{itemize}

\section*{What about  putative narrow N$^*$(1685) in $\eta$ photoproduction off free proton?}

\noindent
It was predicted that the photoexcitation of the charge component of the anti-decuplet nucleon
is strongly suppressed \cite{max}. That makes its search more sophisticated. 
 In Refs.~\cite{acta,jetp} a sharp resonant structure at $W\sim 1685$~MeV was
found in the beam asymmetry data for the $\eta$
photoproduction on the free proton.
\begin{figure}
\vspace*{0.6cm}
\centerline{\epsfverbosetrue\epsfxsize=9cm\epsfysize=5.8cm\epsfbox{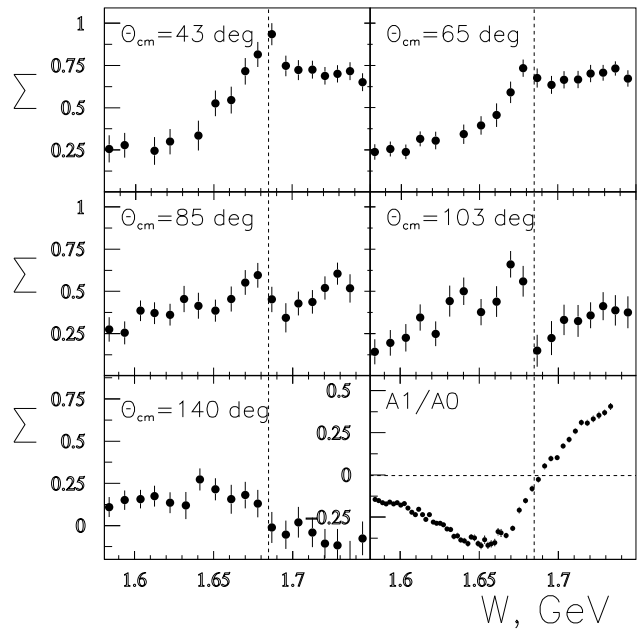}}
\caption{Photon beam asymmetry in $\gamma p\to \eta p$ extracted from the GRAAL
data, see Ref.~\cite{acta,jetp}. The low right figure shows the ratio of Legendre coefficient 
$A_1/A_0$ (\ref{csexp})} extracted from data of Ref.~\cite{Mainz}. \vspace*{-0.3cm}
\label{fig:fr} \vspace{0.3cm}
\end{figure}
Any resonance whose photoexcitation on
the proton is suppressed  may manifest itself in
polarization observables due to interference effects.
The results of Refs.~\cite{acta,jetp} for the beam asymmetry in the $\eta$
photoproduction on the free proton are shown in Fig.~\ref{fig:fr}. One sees that around $W\sim1685$~MeV (shown by 
the vertical dashed line) there is a narrow structure, which looks like a peak at forward angles and which develops 
into an oscillating structure at larger scattering angles. Such behaviour is typical for interference effects of a narrow resonance with smooth
background.

 Fits to the data provided an estimate of the photocoupling
for the charge component of N$^*$(1685)~\cite{acta,jetp}:
\begin{eqnarray}
\label{ap}
\sqrt{{\rm Br}_{\eta N}} A_{1/2}^p \sim 1\cdot 10^{-3}\ {\rm GeV}^{-1/2} .
\end{eqnarray}
One sees that the photocoupling of N$^*$(1685) to the proton is much smaller than the couplng to the neutron (\ref{an}).

Photocouplings (\ref{an}) and (\ref{ap}) correspond to the following resonance cross section at its maximum\footnote{We emphasize that
the theoretical uncertainties in the estimates of the photocouplings (\ref{an}) and (\ref{ap}) are rather large $\pm 40$\%. That can lead to
$\pm 80$\% uncertainties in the estimates of the resonance cross sections.} (at $W=M_R$):

\be
\label{cs}
\sigma_{\rm res}(\gamma n\to \eta n)|_{W=M_R}\sim 8.5\ \left(\frac{10\ {\rm MeV}}{\Gamma_{\rm tot}}\right)\ \mu {\rm b}, \\
\nonumber
\sigma_{\rm res}(\gamma p\to \eta p)|_{W=M_R}\sim 0.04\ \left(\frac{10\ {\rm MeV}}{\Gamma_{\rm tot}}\right)\ \mu {\rm b}.
\ee
Typical values of the non-resonant cross section at $W\sim 1680$~MeV is $\sigma_n\sim 5-6\ \mu$b for the neutron and $\sigma_p\sim 3\ \mu$b
for the proton. One sees from that rough estimate that the resonance cross section on the proton is very small and even in a measurement with an ideal resolution
it is almost impossible to see the corresponding resonance signal. The signal of weak resonance can be revealed through its quantum interference with the strong but smooth
background amplitude, see e.g. \cite{amarian,azimov}. The interference enhancement of a weak signal was used in Refs.~\cite{acta,jetp} to reveal the signal of narrow
N$^*$(1685) in polarization observables. Note that in the case of interference a weak signal can appear not necessarily as a resonance bump but as a dip or
a structure  oscillating with energy.

In order to reveal a weak signal of N$^*$(1685) in the cross section of $\gamma p \to \eta p$ processes one 
needs to perform detailed PWA. Here we just make
a ``back of an envelope" estimate. As we mentioned already a weak resonance should appear as a bump, dip or oscillating structure in the cross section. The {\it maximally possible} magnitude
of such structure can be estimated  as:
\be
\label{est}
\Delta\sigma_{\rm tot}=2\sqrt{\sigma_p\ \sigma_{\rm res}(\gamma p\to \eta p)|_{W=M_R}}\sim 0.7 \ \mu {\rm b},
\ee
that number corresponds to $\sim 0.06$$\ \mu$b/sr in the differential cross section.
Note that the actual magnitude of the interference structure must be smaller than the above value, as the estimate (\ref{est}) assumes that
only one partial wave with quantum numbers of the putative resonance contributes to the cross section. 

Recently the Crystal Ball Collaboration at MAMI published high precision data on $\eta$ photoproduction on free proton \cite{Mainz}.
The cross section was measured with fine steps in the photon energy. The authors of Ref.~\cite{Mainz} concluded that ``
... cross sections for the free proton show no
evidence of enhancement in the region $W \sim 1680$~MeV,
contrary to recent equivalent measurements on the quasifree
neutron. However, this does not exclude the existence
of an N$^*$(1680) state...". As we discussed above one should expect that the 
putative N$^*$(1685) can be seen in the cross section only due to its interference
with strong smooth background and the corresponding signal is not necessarily looks like a peak but rather as the 
structure oscillating with energy  or as a dip.
\begin{figure}
\vspace*{0.6cm}
\centerline{\epsfverbosetrue\epsfxsize=10.1cm\epsfysize=6.5cm\epsfbox{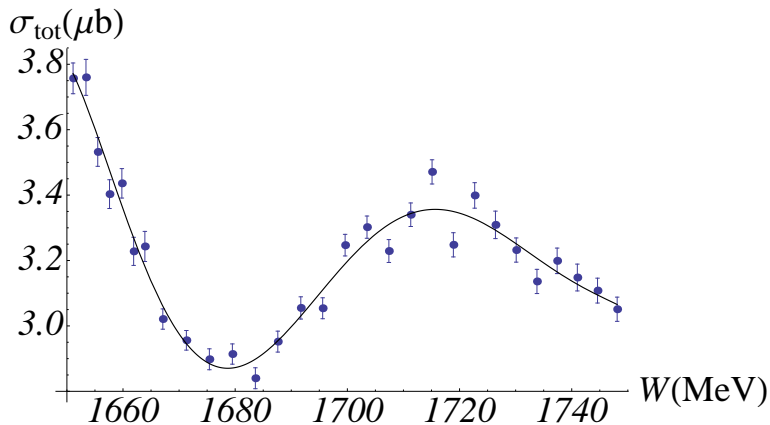}}
\caption{Total cross section of $\gamma p\to \eta p$ process. The data points are from Ref.~\cite{Mainz}. Solid line is the 
6th order polynomial fit to the experimental points just to guide the eye.} \vspace*{-0.3cm}
\label{fig:Stot} \vspace{0.3cm}
\end{figure}

Let us look more carefully at the energy behaviour of the total cross section in the energy region around $W\sim 1685$~MeV. The data of Ref.~\cite{Mainz}
for the total cross section of $\gamma p\to \eta p$ for $W$ in the interval 1650-1750~MeV are shown in Fig.~\ref{fig:Stot}. One sees clearly an oscillation structure
with the distance  between two extrema of  $\Delta W\sim 40$~MeV
(a minimum at $W\sim 1680$~MeV and a maximum at $W\sim 1720$~MeV).
The amplitude of that oscillation structure (the difference between the values of the cross section at the extrema) is about $\sim 0.5\ \mu$b (cf. our ``back of an envelope" estimate
(\ref{est})). We see that in the invariant energy region 1680-1720~MeV
the total cross section of $\gamma p\to \eta p$
reveals a narrow oscillation (or maybe dip) structure with the magnitude compatible with our expectations (\ref{est}) for the interference pattern of the narrow N$^*$(1685)\footnote{We note that this oscillation structure is also seen in the data of Ref.~\cite{crede}, however the authors attributed the structure to an instrumental effect}. 
The amplitude of the oscillation structure and its width are too close to the upper limits what one can expect for the putative
narrow resonance N$^*$(1685). It seems that several partial waves are in play. It might be that the wide resonances 
in the neighbourhood of $W\sim 1685$~MeV, such as $P_{11}(1710)$, $P_{13}(1720)$ and $D_{15}(1675)$ can contribute
additionally to the enhancement of the observed oscillation. All these contributions can be disentangled
by PWA.

\begin{table}[htdp]
\caption{Interference of various partial waves in coefficients $A_i$ (\ref{csexp}). The Legendre coefficient $A_1$
is highlighted because experimentally it clearly exhibits the rapid energy dependence at $W\sim 1650-1750$~MeV. }
\begin{center}
\begin{tabular}{|c|c|c|c|c|c|}
\hline
                & $S_{11}$ & $ P_{11}$ & $P_{13}$ &$D_{13}$ & $D_{15}$\\
\hline
$S_{11}$ & $A_0 $  &  $ {\bf A_1}$  & ${\bf A_1}$      & $A_2$   & $A_2$   \\
\hline
$P_{11}$ & ${\bf A_1}$   &$A_0 $    & $A_2$     & ${\bf A_1}$    &  $A_3$  \\
\hline
$P_{13}$ &  ${\bf A_1}$  & $A_2$    &  $A_0,A_2 $    & ${\bf A_1},A_3$    & ${\bf A_1},A_3$   \\
\hline
$D_{13}$ &  $A_2$  &${\bf A_1}$    &${\bf A_1} ,A_3$      &   $A_0,A_2 $ & $A_2,A_4$   \\
\hline
$D_{15}$ &  $A_2$  &$A_3$    & ${\bf A_1}, A_3$     &$A_2,A_4$    & $A_0,A_2,A_4 $   \\
\hline
\end{tabular}
\end{center}
\label{default}
\end{table}%

The consideration above
shows that around $W\sim 1680$~MeV there exists a phenomenon with the typical energy scale of about 20-40~MeV. In order to  
investigate a possible origin of the phenomenon let us consider the differential cross section. It is convenient to expand the differential cross section
in the Legendre series:
\be
\label{csexp}
\frac{d\sigma}{d\Omega}=\frac{1}{4\pi}\  \sum_{l=0}^\infty A_l(W)\ P_l(\cos\theta),
\ee
where $P_l$ are Legendre polynomials. Note that by definition the coefficient $A_0(W)$ coincides with the total cross section.
The coefficients $A_i(W)$ receive contribution from interference of various partial waves. The partial waves  (for $l\leq 2$) which 
interfere in a given coefficient $A_i(W)$
are listed in Table~1.  As an entry in the table we show the coefficients $A_i$ in which two chosen partial waves interfere.

\begin{figure}
\vspace*{0.6cm}
\centerline{\epsfverbosetrue\epsfxsize=10.1cm\epsfysize=6.5cm\epsfbox{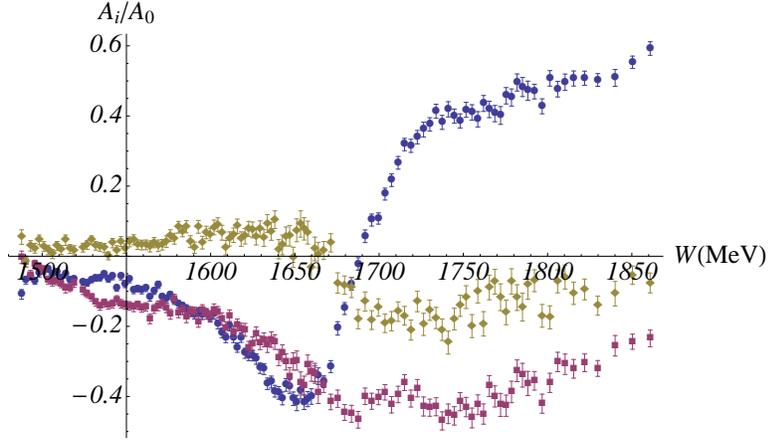}}
\caption{Coefficients $A_i$ of the Legendre expansion (\ref{csexp}) normalized to the total cross section (to $A_0$).
The coefficients $A_i$ are calculated using the data of Ref.~\cite{Mainz}.  The filled circles correspond to $A_1/A_0$, filled squares to $A_2/A_0$
and filled diamods to $A_3/A_0$.
} \vspace*{-0.3cm}
\label{fig:Ai1} \vspace{0.3cm}
\end{figure}

In Fig.~\ref{fig:Ai1} we show the
 normalized Legendre coefficients (\ref{csexp}) ($A_i/A_0$) extracted from the data 
of Ref.~\cite{Mainz}.  One sees that $A_1$ coefficient undergoes rapid change
of its sign on the invariant energy interval of $W\sim 1650-1730$~MeV. Also $A_3$ changes its sign on that interval, whereas
the coefficient $A_2$ shows little structure on that energy interval.  We note that the rapid change of $A_1$ coefficient
occurs exactly at invariant energy where the rapid change of photon beam asymmetry was observed in Refs.~\cite{acta,jetp}.
To illustrate this we show in Fig.~\ref{fig:fr}  the ratio of Legendre coefficient 
$A_1/A_0$ (\ref{csexp}) extracted from data of Ref.~\cite{Mainz} (low right insert) together with photon beam asymmetry
of Refs.~\cite{acta,jetp}.  

It is clear from Table~1 that the rapid change of the
sign of $A_1$ can be driven by the interference of various partial waves. Thus one definitely needs sizable values of
$P$ and/or $D$ waves in the invariant energy interval of $W\sim 1650-1750$~MeV. That simple observation casts serious 
doubts on the model of Ref.~\cite{dor}, which predicts the dominance of $S$-wave in that energy interval.

\begin{figure}
\vspace*{0.6cm}
\centerline{\epsfverbosetrue\epsfxsize=10.1cm\epsfysize=6.5cm\epsfbox{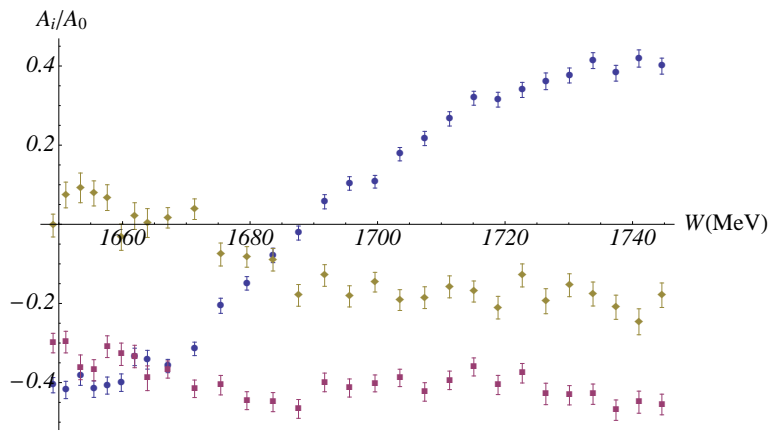}}
\caption{The same as Fig.~\ref{fig:Ai1} for the narrower invariant energy interval of 1650-1750~MeV.
} \vspace*{-0.3cm}
\label{fig:Ai2} \vspace{0.3cm}
\end{figure}
In Fig.~\ref{fig:Ai2} we show normalized coefficients $A_i$ on the narrower energy interval of 1650-1750~MeV.
The Legendre coefficient $A_1$  exhibits rapid change on this small energy interval\footnote{ The coefficient  $A_3$ 
also changes its sign in this energy region, but slower than $A_1$}.
As we discussed above, the total cross section also shows the oscillation structure on the 
energy interval of 1650-1750~MeV (see Fig.~\ref{fig:Stot}). 
The width of the apparently seen structure in $A_1$ is wider than 
in $\sigma_{\rm tot}$ ($\sim 80$~MeV versus $\sim 40$~MeV).
Also the magnitude of the structure is larger than one can expect for the weak contribution of N$^*$(1685). It seems
that other wide resonances contribute to the normalized $A_1$, that can be $P_{11}(1710)$, $P_{13}(1720)$, $D_{15}(1675)$.
These resonances have masses around $W\sim 1685$~MeV and can also (in addition to putative N$^*$(1685)) lead to the
change of the sign of $A_1$. To disentangle the contribution of these resonances one needs detailed PWA, which is beyond
scope of these notes. Here we just make simple estimates to single out the ``rapid" degrees of freedom from the data.  

The main distinctive feature of  putative N$^*$(1685) is its small width, one may try to single out its contribution
to $A_1$ considering derivatives $dA_1/d W$. 
Indeed, looking at  Fig.~\ref{fig:Ai2} one might see that the speed of $A_1$'s change with $W$ has probably a qualitatively different
regime on narrow energy interval of $W\sim 1670-1700$~MeV.  That observation  invites us to study the``speed characteristic" of the normalized $A_1$:
\be
\label{speed}
S_1(W)\equiv W\frac{d}{dW}\left( \frac{A_1(W)}{A_0(W)}\right).
\ee
That quantity is dimesionless, it allows us to separate rapidly changing contributions from  contributions of wide resonances
and smooth background. It is difficult to extract  $S_1(W)$ from the data because of statistical fluctuations in the data that
induce large instabilities in the calculations of the derivative. We use the following procedure to compute $S_1(W)$:
for each $i$th bin in $W$  we choose the energy interval $[W_i, W_{i+12}]$ (about 30~MeV wide) 
and fit the data by the 4th order 
polynomial (13 data points). After that, using resulting from the fit polynomial, we compute $S_1(W)$ analytically  for the 4 middle bins in the interval $[W_i, W_{i+12}]$.  Obviously, the resulting value of $S_1(W)$ for a given $W$ depends on  the initial 
bin in our procedure. The differences of values of $S_1(W)$ reflect the uncertainties in differentiation of the numerical data.  

\begin{figure}
\vspace*{0.6cm}
\centerline{\epsfverbosetrue\epsfxsize=10.1cm\epsfysize=6.5cm\epsfbox{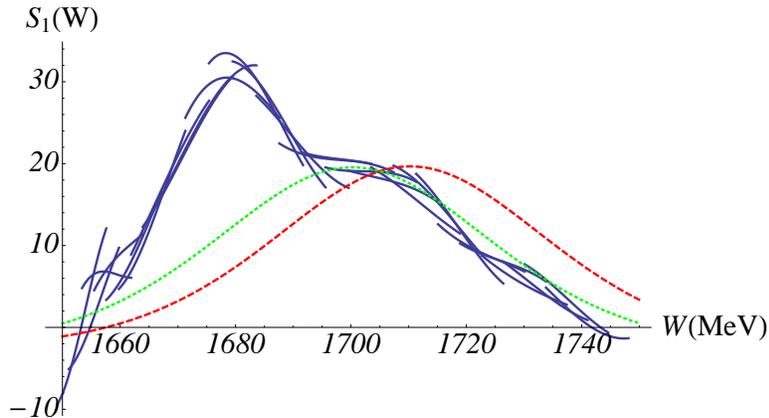}}
\caption{Extracted values of $S_1(W)$ (\ref{speed}) on the energy interval of 1650-1750~MeV. 
As an example,  we show by the dashed line the
contribution of 
100~MeV wide  $P_{11}$ resonance with the mass of 1710~MeV (dotted line corresponds to $M_R=1700$~MeV) and 
$\sqrt{{\rm Br}_{\eta N}} A_{1/2}^p\sim 8\cdot 10^{-3}$~GeV$^{-1/2}$ (that corresponds to $\sigma_{\rm res}/\sigma_{\rm tot}\sim 0.1$). The values of the mass and width are chosen in accordance with the central values for those parameters provided by the Particle Data Group 
\cite{PDG} for the three star N(1710) resonance.  
} \vspace*{-0.3cm}
\label{fig:S1} \vspace{0.3cm}
\end{figure}

In Fig.~\ref{fig:S1} we plot $S_1(W)$ obtained by that procedure. We see that at $W$ around 1660~MeV and 1690~MeV
the ``speed characteristic" $S_1(W)$ (\ref{speed}) is very uncertain (one obtains very different values depending on the starting bin), whereas
between these points the $S_1(W)$ is rather stable. That means that at points 1660~MeV and 1690~MeV the change of the regime of the
$W$ dependence of the normalized $A_1$ happens.  Also it is remarkable that $S_1(W)$ reaches
 its maximum at $W\sim 1680$~MeV (that is corresponds to the inflection point of the normalized $A_1$) which is close to zero of $A_1(W)$ at $W\sim 1685$~MeV.
Such situation is typical for the case when $A_1$ appears as the result of interference
of two partial waves: one is smooth (say $S$-wave) and another is dominated by a resonance (say $P$-wave). Note that the 
value of $S_1(W)$ at maximum at 1680~MeV is rather sizable: $S_1^{\rm max}\sim 30$.

If one uses a simple model, which consist of smooth $S_{11}$ amplitude and a narrow $P_{11}$ resonance 
(mass $M_R$ and total width $\Gamma_R$) on the top of smooth $P_{11}$
background one can derive a simple expression for $S_1^{\rm max}$:
\be
\label{speedmax}
S_1^{\rm max}=4 \frac{M_R}{\Gamma_R}\ \sqrt{\frac{\sigma_{\rm res}}{\sigma_{\rm tot}}}\ \sqrt{1-r}\ (1-2 r) ,
\ee 
where $r$ is the fraction of the $P_{11}$ partial wave in $\sigma_{\rm tot}$ at $W=M_R$ and $\sigma_{\rm res}$
is the resonance cross section. We note that this equation is derived under the assumptions that the resonance is weak, i.e.
$\sigma_{\rm res}\ll \sigma_{\rm tot}$. We consider this limit because otherwise (for $\sigma_{\rm res}\sim \sigma_{\rm tot}$) 
the resonance should be seen in the total cross section as a clean cut peak. 

From Eq.~(\ref{speedmax}) one obtains that for the known wide resonances of width $\Gamma_R\sim 100-200$~MeV
$S_1^{\rm max}\leq (22-11)$ even for optimistically large cross section ratio of $\sigma_{\rm res}/ \sigma_{\rm tot}=0.1$.
As an illustration, the contribution of $P_{11}(1710)$ resonance to $S_1(W)$ is shown by the dashed line in Fig.~\ref{fig:S1}.
For the calculations we used the central values\footnote{The case of $M_R=1700$~MeV is shown by dotted line.} of the 
N(1710) parameters listed by the Particle Data Group \cite{PDG}:
$M_R=1710$~MeV, $\Gamma_R=100$~MeV whereas for the photocoupling  
we took $\sqrt{{\rm Br}_{\eta N}} A_{1/2}^p\sim 8\cdot 10^{-3}$~GeV$^{-1/2}$ which corresponds to the maximal value provided by PDG.
The latter value corresponds to $\sigma_{\rm res}/ \sigma_{\rm tot}\sim 0.1$, if one uses the central values of N(1710) parameters
listed by PDG one obtains the contribution to $S_1(W)$ which is about 10 times smaller than the
one shown by the dashed line on Fig.~\ref{fig:S1}.

We also tried to fit the data on the normalized Legendre coefficient $A_1/A_0$ (see Fig.~\ref{fig:Ai2}) on the energy interval
$W\sim 1650-1750$~MeV by smooth $S_{11}$ partial wave plus N(1710) resonance. One can describe
the data pretty well with $M_R=1685$~MeV and $\Gamma_R=100$~MeV, however the photocoupling comes out
very large $\sqrt{{\rm Br}_{\eta N}} A_{1/2}^p\sim 13\cdot 10^{-3}$~GeV$^{-1/2}$. The latter value corresponds 
to the large resonance cross section of $\sigma_{\rm res}\sim 0.7\ \mu$b, which could be easily seen (but actually not seen) in data on 
$\sigma_{\rm tot}$. Although the mass and width resulting from the fit are not in contradiction
with the very uncertain values provided by the PDG \cite{PDG} for N(1710), the value of  the photocoupling
is far larger than that provided by the PDG. We note that the recent GWU PWA found
no evidence for N(1710) \cite{GWU06}. Other PWA groups \cite{maid,BG} definitely 
require this resonance, however with rather different masses
and with width $\geq 150$~MeV.
\footnote{Unfortunately, it is frequent that results of various PWA groups  are in qualitative contradiction 
with each other. For a non-expert in PWA it is usually very difficult to figure out the physics reasons for that differences.}

The aim of above simple exercises was purely illustrative: it shows that
 it is very difficult to obtain the experimental value of $S_1^{\rm max}\sim 30$
by contribution of known wide resonances if the corresponding resonance cross section is not large. 
For the case of the large resonance cross section the corresponding resonance should be visible as a peak
in the differential cross section.
\begin{figure}
\vspace*{0.6cm}
\centerline{\epsfverbosetrue\epsfxsize=10.1cm\epsfysize=6.5cm\epsfbox{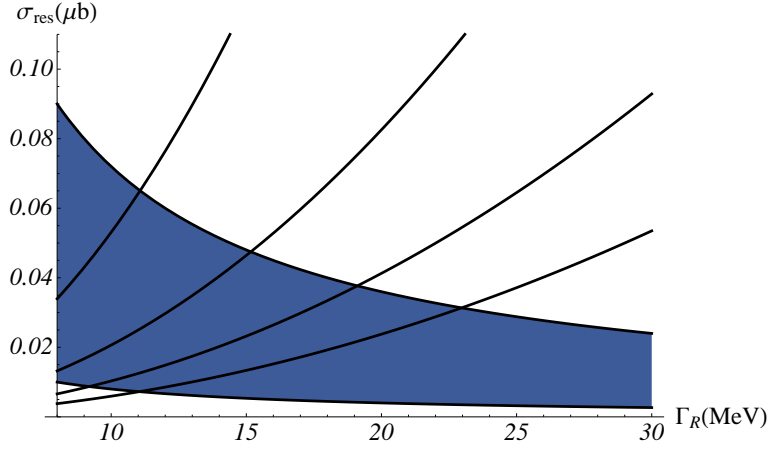}}
\caption{Lines show the relation between the resonance cross section $\sigma_{\rm res}$
and the width of putative resonance $\Gamma_R$ obtained from Eq.~(\ref{speedmax})
with the experimental input $S_1^{\rm max}\sim 30$ and $\sigma_{\rm tot}\sim 3~\mu$b.
The lines correspond to values of the parameter $r=0,0.1,0.2$ and 0.3 (the larger $r$ the steeper the curve).
Shaded area shows our estimate given by Eq.~(\ref{cs}) $\pm 80$\%. 
} \vspace*{-0.3cm}
\label{fig:sg} \vspace{0.3cm}
\end{figure}
According to  Eq.~(\ref{speedmax}), another possibility to obtain the large experimental  value of  
$S_1^{\rm max}\sim 30$ is due to
 the contribution of a narrow resonance with small
photocoupling to the proton (small ratio of cross sections $\sigma_{\rm res}/ \sigma_{\rm tot}$).
From Eq.~(\ref{speedmax}) we see that for each value of parameter $r$  we can determine a relation between the resonance cross
section $\sigma_{\rm res}$ and the resonance total width $\Gamma_R$. 
Taking experimental values of $\sigma_{\rm tot}\sim 3\mu$b
and  $S_1^{\rm max}\sim 30$ we plot  in Fig.~\ref{fig:sg} the relation between the resonance cross section and the resonance width
for several values of the parameter $r$\footnote{To fix this parameter from the experimental data one needs to perform PWA.}.
Also we plot our estimation  of the resonance cross section (\ref{cs}) obtained from the analysis of the beam asymmetry
in $\eta$ photoproduction off free proton \cite{acta,jetp}.  More precisely, we plot the band corresponding to Eq.~(\ref{cs}) $\pm 80$\%
which reflects possible uncertainties in our estimates. For reader's convenience we translated the Fig.~\ref{fig:sg} into the relation 
between the photocoupling $\sqrt{{\rm Br}_{\eta N}} A_{1/2}^p$ and the width of the putative resonance, see Fig.~\ref{fig:ag}.
\begin{figure}
\vspace*{0.6cm}
\centerline{\epsfverbosetrue\epsfxsize=10.1cm\epsfysize=6.5cm\epsfbox{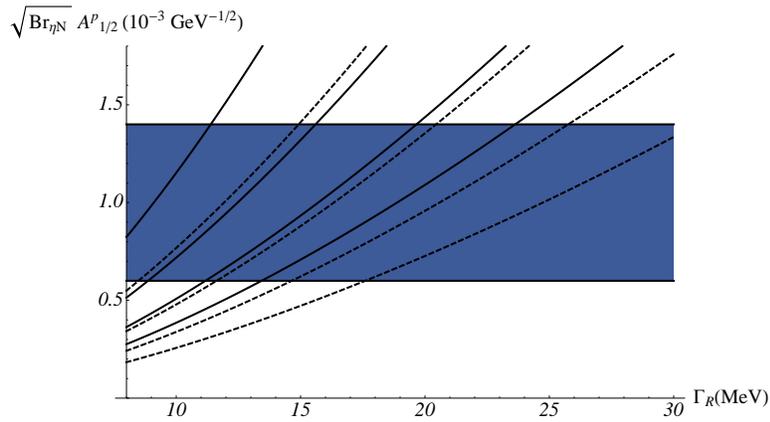}}
\caption{The same as Fig.~\ref{fig:sg}, but translated to the relation between $\sqrt{{\rm Br}_{\eta N}} A_{1/2}^p$ and $\Gamma_R$.
By dashed lines we show the solutions of Eq.~(\ref{speedmax}) for the case of $S_1^{\rm max}\sim 20$.
} \vspace*{-0.3cm}
\label{fig:ag} \vspace{0.3cm}
\end{figure} 
Additionally, in Fig.~\ref{fig:ag} we show  the solutions of Eq.~(\ref{speedmax}) for the case of $S_1^{\rm}\sim 20$ by the dashed lines.
That case takes into account possible contributions of wide resonances to $S_1(W)$ , that resonances can contribute to some part
of experimental value of  $S_1^{\rm}\sim 30$, see e.g. the dashed line in Fig.~\ref{fig:S1}.

Given that our estimates are very rough, the agreement is rather impressive. We can conclude from the presented simple analysis
that the observed in Ref.~\cite{Mainz} oscillation of $\sigma_{\rm tot}(\gamma p\to\eta p)$ and rapid change of the Legendre
coefficient $A_1(W)$ around $W\sim 1685$~MeV may indicate an existence of new narrow N$^*(1685)$ resonance
with $\Gamma_{\rm tot}\leq 50 $~MeV and small resonance photocoupling  in the range of $\sqrt{{\rm Br}_{\eta N}} A_{1/2}^p 
\sim (0.3-3)\cdot 10^{-3}$~GeV$^{-1/2}$. 

\section*{Conclusions}

Recent high precision measurements of the $\gamma p\to\eta p$ cross section \cite{Mainz} 
show the oscillation of $\sigma_{\rm tot}(\gamma p\to\eta p)$ and rapid change of the Legendre
coefficient $A_1(W)$ around $W\sim 1685$~MeV. 
These phenomena occurs at the same energy interval as previously observed in  Refs.~\cite{acta,jetp}
resonance behaviour of the photon beam asymmetry, see Fig.~\ref{fig:fr}.
We made very simple analysis of that phenomena using ``speed characteristics" (\ref{speed}) in order to single out
 ``rapid" contributions on the background of smooth contributions of known wide resonances.
Our analysis showed that the data of \cite{Mainz} may indicate an existence of new narrow N$^*(1685)$ resonance
with $\Gamma_{\rm tot}\leq 50 $~MeV and small resonance photocoupling  in the range of $\sqrt{{\rm Br}_{\eta N}} A_{1/2}^p 
\sim (0.3-3)\cdot 10^{-3}$~GeV$^{-1/2}$.  These parameters are in agreement with the analysis of 
the photon beam asymmetry in $\gamma p\to \eta p$ process performed in Refs.~\cite{acta,jetp}.

The estimates presented here provide us the feeling of the expected scales for the effect of putative N$^*(1685)$
in the cross section of $\gamma p\to\eta p$. The estimates also show that the effect of putative N$^*(1685)$ is
interlaced with effects of  neighbourhood wide resonances, such as $P_{11}(1710)$, $P_{13}(1720)$ and $D_{15}(1675)$. For example, the rapid change of the Legendre coefficient $A_1$ (but not oscillation structure
in $\sigma_{\rm tot}$) can be in principle
described by the contribution of the $100$~MeV wide N(1710) resonance, however its photocoupling
should be unrealistically large $\sqrt{{\rm Br}_{\eta N}} A_{1/2}^p \sim 13\cdot 10^{-3}$~GeV$^{-1/2}$.
Surely, for more detailed separation of the putative narrow N$^*(1685)$ from other contributions one needs 
full PWA. We hope that our simple estimates were able to grasp main physics in observed phenomena and
future PWA will be able to detail our observations.   

It seems that all experimental facts discussed here strongly support the existence of new narrow nucleon excitation 
N$^*(1685)$ with properties\footnote{The phenomenological properties of putative N$^*(1685)$ are summarized 
 concisely
in Ref.~\cite{jetp}} neatly coinciding with those predicted 
for the non-strange member of exotic anti-decuplet
\cite{dia,dia1,arndt,michal,max} (for the most recent analysis of the properties of anti-decuplet baryons see Ref.~\cite{sigma}).

\section*{Acknowledgements}
This work has been
supported in parts by SFB/Transregio~16 (Germany) and  by Basic Science Research Program through the
National Research Foundation of Korea(NRF) funded by the Ministry of
Education, Science and Technology (grant 2010-0013430). 
We are thankful to  M.~D\"oring,  A.~Fix, A.~Sarantsev, I.~Strakovsky and L.~Tiator for interesting discussions and
for correspondance.  MVP is thankful to I.~Strakovsky for communicating data published in Ref.~\cite{Mainz}.

%%%%%%%%%%%%%%%%%%%%%%%%%%%%%%%%%%%%%%%%%%%%%%%%%%%%%%


\begin{thebibliography}{99}


\bibitem{dia}
 D.~Diakonov, V.~Petrov and M.~V.~Polyakov,
  %``Exotic anti-decuplet of baryons: Prediction from chiral solitons,''
  Z.\ Phys.\  A {\bf 359} (1997) 305
  [arXiv:hep-ph/9703373].
  %%CITATION = ZEPYA,A359,305;%%

  \bibitem{dia1}D.~Diakonov and V.~Petrov,
  %``Where are the missing members of the baryon antidecuplet?,''
  Phys.\ Rev.\  D {\bf 69} (2004) 094011
  [arXiv:hep-ph/0310212].
  %%CITATION = PHRVA,D69,094011;%%
\bibitem{arndt} R.~A.~Arndt {\it et al.},
  %``Nonstrange and other unitarity partners of the exotic Theta+ baryon,''
  Phys.\ Rev.\  C {\bf 69}, 035208 (2004)
  [arXiv:nucl-th/0312126].
  %%CITATION = PHRVA,C69,035208;%%

  
\bibitem{michal}
  J.~R.~Ellis, M.~Karliner and M.~Praszalowicz,
  %``Chiral-soliton predictions for exotic baryons,''
  JHEP {\bf 0405} (2004) 002
  [arXiv:hep-ph/0401127];\\
  %%CITATION = JHEPA,0405,002;%%
%\cite{Praszalowicz:2004dn}
  M.~Praszalowicz,
  %``SU(3) breaking in decays of exotic baryons,''
  Acta Phys.\ Polon.\  B {\bf 35} (2004) 1625
  [arXiv:hep-ph/0402038].
  %%CITATION = APPOA,B35,1625;%%
  
\bibitem{max}M.~V.~Polyakov and A.~Rathke,
  %``On photoexcitation of baryon antidecuplet,''
  Eur.\ Phys.\ J.\  A {\bf 18}, 691 (2003)
  [arXiv:hep-ph/0303138].
  %%CITATION = EPHJA,A18,691;%%


\bibitem{gra0}
 V.~Kuznetsov  [GRAAL Collaboration],
  %``eta photoproduction off the neutron at GRAAL: Evidence for a resonant
  %structure at W = 1.67-GeV,''
  arXiv:hep-ex/0409032.
  %%CITATION = HEP-EX/0409032;%%

\bibitem{gra1}  V.~Kuznetsov {\it et al.},
  %``Evidence for a narrow structure at W approx. 1.68-GeV in eta
  %photoproduction on the neutron,''
  Phys.\ Lett.\  B {\bf 647}, 23 (2007)
  [arXiv:hep-ex/0606065].
  %%CITATION = PHLTA,B647,23;%%

\bibitem{kru}  I.~Jaegle {\it et al.}  [CBELSA Collaboration and TAPS Collaboration],
  %``Quasi-free photoproduction of eta-mesons of the neutron,''
  Phys.\ Rev.\ Lett.\  {\bf 100} (2008) 252002
  [arXiv:0804.4841 [nucl-ex]].
  %%CITATION = PRLTA,100,252002;%%


\bibitem{kas}  F.~Miyahara {\it et al.},
  %``Narrow Resonance At E(Gamma) = 1020-Mev In The D (Gamma, Eta) P N
  %Reaction,''
  Prog.\ Theor.\ Phys.\ Suppl.\  {\bf 168}, 90 (2007).
  %%CITATION = PTPSA,168,90;%%

\bibitem{wert}  D.~Werthmuller  [for the Crystal Ball/TAPS collaborations],
  %``Investigation of the anomaly in eta-photoproduction off the neutron,''
  Chin.\ Phys.\  C {\bf 33}, 1345 (2009)
  [arXiv:1001.3840 [nucl-ex]].
  %%CITATION = CHPHD,C33,1345;%%

\bibitem{acta}   V.~Kuznetsov {\it et al.},
  %``Evidence for a narrow N(1685) resonance in eta photoproduction off the
  %nucleon,''
  Acta Phys.\ Polon.\  B {\bf 39}, 1949 (2008)
  [arXiv:0807.2316 [hep-ex]];\\
  %%CITATION = APPOA,B39,1949;%%
  V.~Kuznetsov, {\it et al.},
  %``eta photoproduction on the proton revisited: Evidence for a narrow N(1685)
  %resonance?,''
  arXiv:hep-ex/0703003.
  %%CITATION = HEP-EX/0703003;%%

\bibitem{jetp}  V.~Kuznetsov and M.~V.~Polyakov,
  %``New Narrow Nucleon N*(1685),''
  JETP Lett.\  {\bf 88}, 347 (2008)
  [arXiv:0807.3217 [hep-ph]].
  %%CITATION = JTPLA,88,347;%%

\bibitem{crede}
 V.~Crede {\it et al.}  [CBELSA/TAPS Collaboration],
  %``Photoproduction of eta and eta-prime mesons off protons,''
  Phys.\ Rev.\  C {\bf 80}, 055202 (2009)
  [arXiv:0909.1248 [nucl-ex]].
   %%CITATION = PHRVA,C80,055202;%%
\bibitem{cr2}
F.~Renard {\it et al.}  [GRAAL Collaboration],
  %``Differential cross-section measurement of eta photoproduction on the
  %proton from threshold to 1100-MeV,''
  Phys.\ Lett.\  B {\bf 528} (2002) 215
  [arXiv:hep-ex/0011098];\\
  %%CITATION = PHLTA,B528,215;%%
  M.~Dugger {\it et al.}  [CLAS Collaboration],
  %``Eta photoproduction on the proton for photon energies from 0.75-GeV to
  %1.95-GeV,''
  Phys.\ Rev.\ Lett.\  {\bf 89} (2002) 222002
  [Erratum-ibid.\  {\bf 89} (2002) 249904].
  %%CITATION = PRLTA,89,222002;%%


\bibitem{az}    Y.~I.~Azimov, {\it et al.},
  %``Extraction of radiative decay width for the non-strange partner of
  %Theta+,''
  Eur.\ Phys.\ J.\  A {\bf 25}, 325 (2005)
  [arXiv:hep-ph/0506236].
  %%CITATION = EPHJA,A25,325;%%

\bibitem{kim}  K.~S.~Choi, S.~i.~Nam, A.~Hosaka and H.~C.~Kim,
  %``A new N*(1675) resonance in the gamma N --> eta N reaction,''
  Phys.\ Lett.\  B {\bf 636}, 253 (2006)
  [arXiv:hep-ph/0512136].
  %%CITATION = PHLTA,B636,253;%%


\bibitem{tia}   A.~Fix, L.~Tiator and M.~V.~Polyakov,
  %``Photoproduction of eta-mesons on the deuteron above S11(1535) in the
  %presence of a narrow P11(1670) resonance,''
  Eur.\ Phys.\ J.\  A {\bf 32}, 311 (2007)
  [arXiv:nucl-th/0702034].
  %%CITATION = EPHJA,A32,311;%%



\bibitem{skl}  V.~Shklyar, H.~Lenske and U.~Mosel,
  %``eta-photoproduction in the resonance energy region,''
  Phys.\ Lett.\  B {\bf 650}, 172 (2007)
  [arXiv:nucl-th/0611036].
  %%CITATION = PHLTA,B650,172;%%
\bibitem{ani} A.~V.~Anisovich {\it et al.},
  %``Photoproduction of $\eta$ mesons off neutrons from a deuteron target,''
  Eur.\ Phys.\ J.\  A {\bf 41}, 13 (2009)
  [arXiv:0809.3340 [hep-ph]].
  %%CITATION = EPHJA,A41,13;%%


\bibitem{dor}
M.~Doring and K.~Nakayama,
  %``On the cross section ratio sigma_n/sigma_p in eta photoproduction,''
  Phys.\ Lett.\  B {\bf 683}, 145 (2010)
  [arXiv:0909.3538 [nucl-th]].
  %%CITATION = PHLTA,B683,145;%%


\bibitem{Compton}
  V.~Kuznetsov {\it et al.},
  %``Evidence for Narrow N*(1685) Resonance in Quasifree Compton Scattering on
  %the Neutron,''
  Phys.\ Rev.\  C {\bf 83} (2011) 022201(R)
  [arXiv:1003.4585 [hep-ex]].
  %%CITATION = PHRVA,C83,022201;%%

\bibitem{witthauer}
L.~Witthauer, Master Theses, Basel University (2010),\\
http://jazz.physik.unibas.ch/site/theses.html


\bibitem{krusche}
 Talk of B.~Krusche at the 11th International Workshop on Meson Production, Properties and Interaction,
 Krakow, June 10-15, 2010, see http://meson.if.uj.edu.pl/

 \bibitem{amarian}
 M.~Amarian, D.~Diakonov and M.~V.~Polyakov,
  %``To see the exotic Theta^+ baryon from interference,''
  Phys.\ Rev.\  D {\bf 78} (2008) 074003
  [arXiv:hep-ph/0612150].
  %%CITATION = PHRVA,D78,074003;%%

 \bibitem{azimov}
Y.~Azimov,
  %``Quantum interference of particles and resonances,''
  J.\ Phys.\ G {\bf 37} (2010) 023001
  [arXiv:0904.1376 [hep-ph]].
  %%CITATION = JPHGB,G37,023001;%%


\bibitem{Mainz}
 E.~F.~McNicoll {\it et al.}  [Crystal Ball Collaboration at MAMI],
  %``Study of the gp-->etap reaction with the Crystal Ball detector at the Mainz
  %Microtron(MAMI-C),''
  Phys.\ Rev.\  C {\bf 82} (2010) 035208
  [arXiv:1007.0777 [nucl-ex]].
  %%CITATION = PHRVA,C82,035208;%%

\bibitem{PDG}
  K.~Nakamura {\it et al.}  [Particle Data Group],
  %``Review of particle physics,''
  J.\ Phys.\ G {\bf 37} (2010) 075021.
  %%CITATION = JPHGB,G37,075021;%%

\bibitem{GWU06}
  R.~A.~Arndt, W.~J.~Briscoe, I.~I.~Strakovsky and R.~L.~Workman,
  %``Extended Partial-Wave Analysis of piN Scattering Data,''
  Phys.\ Rev.\  C {\bf 74} (2006) 045205
  [arXiv:nucl-th/0605082].
  %%CITATION = PHRVA,C74,045205;%%

\bibitem{maid}
W.~T.~Chiang, S.~N.~Yang, L.~Tiator, M.~Vanderhaeghen and D.~Drechsel,
  %``A reggeized model for eta and eta' photoproduction,''
  Phys.\ Rev.\  C {\bf 68} (2003) 045202
  [arXiv:nucl-th/0212106];\\
  %%CITATION = PHRVA,C68,045202;%%
A.~Fix and L.~Tiator, private communication.
\bibitem{BG}
  A.~V.~Anisovich, E.~Klempt, V.~A.~Nikonov, A.~V.~Sarantsev and U.~Thoma,
  %``P-wave excited baryons from pion- and photo-induced hyperon production,''
  arXiv:1009.4803 [hep-ph].
  %%CITATION = ARXIV:1009.4803;%%


\bibitem{sigma}
 K.~Goeke, M.~V.~Polyakov and M.~Praszalowicz,
  %``On strange SU(3) partners of Theta+,''
  Acta Phys.\ Polon.\  B {\bf 42} (2011) 61
  [arXiv:0912.0469 [hep-ph]];\\
  %%CITATION = APPOA,B42,61;%%
   M.~Praszalowicz,
  %``Importance of Mixing for Exotic Baryons,''
  Acta Phys.\ Polon.\ Supp.\  {\bf 3} (2010) 917
  [arXiv:1005.1007 [hep-ph]].
  %%CITATION = APPXA,3,917;%%


\end{thebibliography}
\end{document}